\documentclass[aps,prx,twocolumn,notitlepage,groupedaddress,showpacs,longbibliography]{revtex4-1}

\usepackage{amsmath}    
\usepackage{amsfonts}
\usepackage{bm}
\usepackage{amssymb}
\usepackage{appendix}
\usepackage{graphicx}   
\usepackage{color}      
\usepackage{subfigure}  
\usepackage{comment}

\newcommand{\re}{\textrm{Re}}
\newcommand{\im}{\textrm{Im}}

\begin{document}

\title{Effects of non-Hermitian perturbations on Weyl Hamiltonians with arbitrary topological charges}
\author{Alexander Cerjan, Meng Xiao, Luqi Yuan, and Shanhui Fan} 
\affiliation{Department of Electrical Engineering, and Ginzton Laboratory, Stanford  University,  Stanford,  California  94305,  USA}

\date{\today}

\begin{abstract}
  We provide a systematic study of non-Hermitian topologically charged systems.
  Starting from a Hermitian Hamiltonian supporting Weyl points with arbitrary topological charge,
  adding a non-Hermitian perturbation transforms the Weyl points to one-dimensional exceptional contours.
  We analytical prove that the topological charge is preserved on the exceptional contours. 
  In contrast to Hermitian systems, the addition of gain and loss allows for a new class of topological phase
  transition: when two oppositely charged
  exceptional contours touch, the topological charge can dissipate without opening a gap. These effects can be demonstrated in
  realistic photonics and acoustics systems.
\end{abstract}

\maketitle

\section{Introduction}

The study of topological systems represents an important frontier in both
condensed matter physics and photonics, as these systems can possess exotic electronic and photonic states, 
such as chiral surface states which realize non-reciprocal transport \cite{kane_quantum_2005,konig_quantum_2007,haldane_possible_2008,hsieh_topological_2008,raghu_analogs_2008,wang_observation_2009,koch_time-reversal-symmetry_2010,umucalilar_artificial_2011,hafezi_robust_2011,fang_realizing_2012,kraus_topological_2012,kitagawa_observation_2012,rechtsman_photonic_2013,khanikaev_photonic_2013,hafezi_imaging_2013}.
A defining feature of these unusual topological states is their protection against many forms
of disorder.

One important class of three-dimensional topological systems are 
Weyl semi-metals \cite{wan_topological_2011,yang_quantum_2011,lu_weyl_2013,xu_discovery_2015,lv_observation_2015,yang_weyl_2015,soluyanov_type-ii_2015,xiao_synthetic_2015,lu_symmetry-protected_2016,chen_photonic_2016,lin_photonic_2016,xiao_hyperbolic_2016,gao_photonic_2016,fang_topological_2016,noh_experimental_2017}, which possess a set of isolated degeneracies in their
band structure. These degeneracies possess topological charges and represent sources or sinks of Berry flux \cite{berry_quantal_1984}.
It is known that in a system with Weyl points, any Hermitian perturbation can only change the location of the Weyl
points and can not remove or create them. Such Hermitian perturbations are common in electronic systems.
On the other hand, in photonic systems there are many perturbations, such
as material gain and absorption, as well as radiative outcoupling, which break Hermiticity. Moreover, many recent
studies have indicated that introducing non-Hermitian perturbations in topologically trivial
systems can result in unusual phenomena, such as promoting single mode operation in lasers \cite{bachelard_taming_2012,hisch_pump-controlled_2013,hodaei_parity-time_symmetric_2014,feng_single-mode_2014,liew_active_2014,liew_pump-controlled_2015,cerjan_controlling_2016},
loss-induced transmission in waveguide arrays \cite{guo_observation_2009,cerjan_eigenvalue_2016}, reverse pump dependence in lasers \cite{liertzer_pump-induced_2012,brandstetter_reversing_2014,peng_loss-induced_2014},
and control over pairs of polarization states \cite{lawrence_manifestation_2014,cerjan_achieving_2017}.
Analogously, therefore, it is important to understand how non-Hermitian perturbations influence
the properties of topologically non-trivial systems \cite{esaki_edge_2011,hu_absence_2011,malzard_topologically_2015,lee_anomalous_2016,leykam_edge_2017,xu_weyl_2017,hu_exceptional_2017,weimann_topologically_2017,shen_topological_2017}. 

In previous
studies of topologically trivial periodic systems, it was discovered that when a non-Hermitian perturbation is introduced, 
degeneracies in the band structure of the underlying Hermitian system can transform to 
rings of exceptional points where both the eigenvalues and eigenvectors become
identical, and the system has a non-trivial Jordan normal form \cite{bender_pt-symmetric_1999,bender_complex_2002,makris_beam_2008,szameit_pt-symmetry_2011,zhen_spawning_2015,cerjan_zipping_2016,mock_parity-timesymmetry_2016,cerjan_effects_2016,mock_characterization_2016}. 
More recently, it was discovered that starting from a Hermitian system supporting
charge-$1$ Weyl points, which is a topologically non-trivial degeneracy, introducing a non-Hermitian perturbation can transform a Weyl point into an exceptional ring,
with the Berry charge of the original Weyl point preserved on the ring \cite{xu_weyl_2017,shen_topological_2017}.
Integrating the Berry curvature on a surface surrounding the exceptional ring yields a quantized
Berry charge, while integrating the Berry curvature on a surface inside the exceptional ring
yields no charge. All of these studies point to the connection between degeneracies
in Hermitian systems, and the creation of exceptional rings when such systems
are subject to non-Hermitian perturbations. However, there has not been a general
treatment of the effects of breaking the Hermiticity of Weyl points with an arbitrary charge,
nor a systematic approach to the unusual properties of these exotic systems.

In this paper, we provide a systematic study of non-Hermitian topologically charged systems.
Our study uncovers a set of remarkable effects in this class of systems.
First, we analytically prove that in the presence of arbitrary non-Hermitian perturbation,
a Weyl point with an arbitrary charge also transforms into a closed one-dimensional exceptional 
contour, with the topological charge preserved on the contour. However, such a contour need not
form a single ring, but can take a more complex shape when the charge is greater than one. Second, 
we demonstrate that, in contrast to Hermitian systems, the addition of gain and loss allows for an alternative mechanism by
which the topological charge in the system can dissipate: when two oppositely charged
exceptional contours touch, the resulting exceptional contour does not possess a Berry charge.
Unlike a Hermitian system, here the disappearance of the charge is not associated with the opening of the band gap.
Third, in such systems
the upper and lower bands associated with the exceptional ring are two branches of the same Riemann sheet, and so it
is possible to follow a smooth path through the exceptional contour and transition from being
on the upper band to being on the lower band. Finally, all of these effects
can be demonstrated in realistic photonic and acoustic systems.

The remainder of this paper is organized as follows. In Sec.\ \ref{sec:theory}, we present
a general theoretical treatment using an effective two-band Hamitonian. In particular, Sec.\ \ref{sec:2b} provides
the analytic proof of the conservation of topological charge on the exceptional contour,
and Sec.\ \ref{sec:2d} describes how the addition of gain and
loss can dissipate topological charge. In Sec.\ \ref{sec:3} we observe a pair of charge-2 Weyl
points which transform into non-ring exceptional contours in a tight binding model of a photonic system.
In Sec.\ \ref{sec:haldane} we discuss the effects of gain and loss on the chiral edge modes of
a system. Finally, we offer some concluding remarks in Sec.\ \ref{sec:4}.

\section{Models of non-Hermitian topological systems \label{sec:theory}}

\subsection{Formation of exceptional contours from charge-$n$ Weyl points \label{sec:2a}}

One of the most important differences between Hermitian and non-Hermitian systems are the
types of eigenvalue degeneracies that each system type displays. Although both systems can
display ordinary degeneracies where two or more eigenvalues become equal, non-Hermitian systems
can also possess exceptional points, where not only are the eigenvalues equal, but the eigenvectors
become identical and self-orthogonal, and the system has a non-trivial Jordan normal form \cite{kato_perturbation_1995,heiss_exceptional_2004,heiss_physics_2012}.

In spite of the differences, there is in fact a general connection between the band degeneracies
in Hermitian systems, and the exceptional points when a non-Hermitian perturbation is added to such
systems \cite{ge_parity-time_2014,zhen_spawning_2015,cerjan_zipping_2016}. To illustrate this connection in topologically non-trivial systems,
we first consider a general three-dimensional Hermitian system with a charge-$n$ Weyl point.
The $2 \times 2$ Hamiltonian in the vicinity of such a Weyl point has the form \cite{fang_multi-weyl_2012}, 
\begin{equation}
  H(\mathbf{k}) = k_+^{n} \sigma_+ + k_-^{n} \sigma_- + k_z \sigma_z + \omega_0 I, \label{eq:Hherm}
\end{equation}
in which $k_\pm = (k_x \pm ik_y)$, $\sigma_\pm = 1/2(\sigma_x \pm i\sigma_y)$,
$\sigma_{x,y,z}$ are the Pauli matrices, $I$ is the $2 \times 2$ identity matrix, and $\omega_0$ is the frequency of
the Weyl point. In the vicinity of the Weyl point the two bands are described by
\begin{equation}
  \lambda_\pm = \omega_0 \pm \sqrt{(k_x^2 + k_y^2)^n + k_z^2}.
\end{equation}
To achieve a degeneracy for which $\lambda_+=\lambda_-$, we must have $k_x = k_y = k_z = 0$. This is an
example of the general result that in a Hermitian system, three constraints must be simultaneously satisfied to achieve an
accidental degeneracy in a system, i.e.\ a degeneracy that is not protected by a symmetry in the system \cite{von1929some,berry_quantal_1984}.
As a result, for a Hermitian system in three dimensional space, absent of symmetry, degeneracy can only be found at isolated points in $k$-space. 

\begin{figure}[t!]
  \centering
  \includegraphics[width=0.98\linewidth]{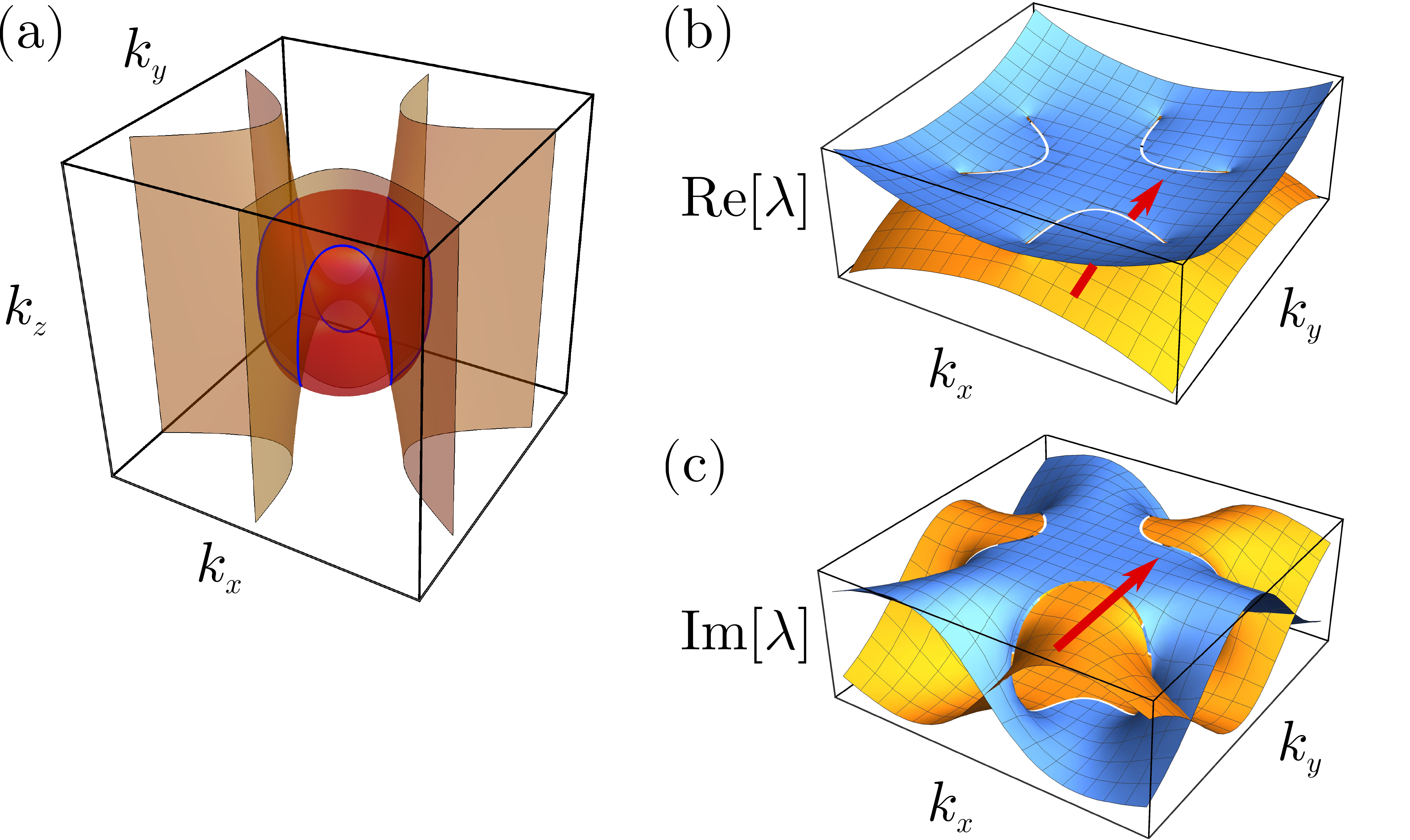}
  \caption{(a) Exceptional contour associated with a charge-3 Weyl point under non-Hermitian perturbation, as described by Eq.\ (\ref{eq:nhw}) with $n=3$ and
    $\boldsymbol{\tau} = (1,0,0.3)$. The two surfaces plotted are Eq.\ (\ref{eq:5}) (red) and Eq.\ (\ref{eq:6}) (orange),
    and their intersection defines the exceptional contour (blue). (b) and (c) Real and imaginary parts of $\lambda_\pm$ are shown
    in blue and yellow, respectively. The red arrow indicates a smooth trajectory across the corresponding Riemann surface,
    it starts on the lower branch and ends on the upper branch.}
  \label{fig:0}
\end{figure}

We now add a generic non-Hermitian term to the previous Hamiltonian from Eq.\ (\ref{eq:Hherm}), so that
\begin{equation}
  H(\mathbf{k}) = k_+^{n} \sigma_+ + k_-^{n} \sigma_- + k_z \sigma_z + \omega_0 I + i \boldsymbol{\tau} \cdot \boldsymbol{\sigma}, \label{eq:nhw}
\end{equation}
in which $\boldsymbol{\tau} = (\tau_x, \tau_y, \tau_z) \in \mathbb{R}$, $\boldsymbol{\sigma} = (\sigma_x, \sigma_y, \sigma_z)$,
and we now allow $\omega_0$ to be complex. Equation (\ref{eq:nhw}) yields a band structure of
\begin{multline}
  \lambda_\pm(\mathbf{k}) = \omega_0 \pm \left[k_\rho^{2n} + k_z^2 - \boldsymbol{\tau}^2 \right. \\
    \left. + 2i(\tau_x k_\rho^n \cos(n \phi) - \tau_y k_\rho^n \sin(n \phi) + \tau_z k_z) \right]^{1/2}, \label{eq:lam1}
\end{multline}
in which we have adopted cylindrical coordinates, $(k_\rho, \phi, k_z)$, with $k_\pm = k_\rho e^{\pm i \phi}$.
In contrast to the Hermitian system, there are now only two criteria which must be met to find
$\lambda_+ = \lambda_-$,
\begin{align}
  0 =& \; (k_x^2 + k_y^2)^n + k_z^2 - \boldsymbol{\tau}^2, \label{eq:5} \\
  0 =& \; \tau_x k_\rho^n \cos(n \phi) - \tau_y k_\rho^n \sin(n \phi) + \tau_z k_z. \label{eq:6}
\end{align}
Equations (\ref{eq:5}) and (\ref{eq:6}) define a single closed contour of degeneracies in $\mathbf{k}$-space.
To prove that this contour consists entirely of exceptional points, we calculate the inner product
of the right eigenvectors, $H|\psi_\pm^R \rangle = \lambda_\pm |\psi_\pm^R \rangle$, and left eigenvectors,
$\langle \psi_\pm^L | H = \lambda_\pm \langle \psi_\pm^L |$,
\begin{equation}
  \langle \psi_\pm^L | \psi_\pm^R \rangle = 2(\lambda_\pm -\omega_0) \left(\lambda_\pm -\omega_0 + k_z + i \tau_z \right),
\end{equation}
which is necessarily zero on this degenerate contour, demonstrating that the eigenstates become self-orthogonal,
one of the signatures of being at an exceptional point \cite{kato_perturbation_1995,heiss_exceptional_2004,heiss_physics_2012}.
Therefore, we see that a topologically charged point degeneracy in a Hermitian system gives
rise to a contour of exceptional points when a non-Hermitian perturbation is added. An example
of an exceptional contour existing at the intersection of the two surfaces given in Eqs.\ (\ref{eq:5}) and (\ref{eq:6})
is shown in Fig.\ \ref{fig:0}(a) for $n=3$.

As $\lambda_\pm(\mathbf{k})$ is a multi-valued complex function in non-Hermitian systems, the $\pm$ signs in Eq.\ (\ref{eq:lam1}) identify the two
distinct branches of this function. Alternatively, one can view the two branches as being a part
of the same Riemann surface, and rewrite Eq.\ (\ref{eq:lam1}) as
\begin{equation}
  \lambda_\theta(\mathbf{k}) = \omega_0 + \sqrt{a(\mathbf{k})}e^{i \theta/2}, \label{eq:lam2}
\end{equation}
in which
\begin{align}
  \cos(\theta) =& \frac{k_\rho^{2n} + k_z^2 - \boldsymbol{\tau}^2}{a(\mathbf{k})}, \label{eq:cos} \\
  \sin(\theta) =& \frac{2(\tau_x k_\rho^n \cos(n \phi) - \tau_y k_\rho^n \sin(n \phi) + \tau_z k_z)}{a(\mathbf{k})}, \label{eq:sin} \\
  a(\mathbf{k}) =& \left[\left(k_\rho^{2n} + k_z^2 - \boldsymbol{\tau}^2 \right)^2 \right. \notag \\
    & \left. +  4(\tau_x k_\rho^n \cos(n \phi) - \tau_y k_\rho^n \sin(n \phi) + \tau_z k_z)^2 \right]^{1/2},
\end{align}
so that $\lambda_\theta(\mathbf{k})$ only returns to its original value for $\theta \rightarrow \theta + 4\pi$.
The fact that the two bands constitute
different branches of the same Riemann surface has a physical consequence, if one starts with a state
on the lower band and travels through the exceptional contour where the upper and lower branches meet,
the state smoothly transitions to being on the upper band. An example of this process can be seen in
Fig.\ \ref{fig:0}(b) and (c), where the red arrow marks a smooth trajectory across the Riemann surface which
exists on both branches of $\lambda_\pm$. As we will demonstrate in Sec.\ \ref{sec:haldane},
this effect can also be seen in the projection of the bulk bands of semi-infinite systems.

\subsection{Topological charge of an exceptional contour \label{sec:2b}}

To define a topological charge for the exceptional contour, the Berry connection, $\mathbf{A}(\mathbf{k})$, and Berry curvature, $\boldsymbol{\Omega}(\mathbf{k})$, must
be generalized to non-Hermitian systems, for which the left and right eigenvectors
are not necessarily related by the conjugate transpose, i.e.\ $\langle \psi^L | \ne (|\psi^R \rangle)^\dagger$. Although there are four
reasonable possibilities for generalizing the Berry connection which correspond to using different combinations of the left
and right eigenstates in the definition,
\begin{equation}
  \mathbf{A}^{(L/R),(L/R)} = i \langle \psi(\mathbf{k})^{(L/R)} | \nabla_{\mathbf{k}} | \psi(\mathbf{k})^{(L/R)} \rangle,
\end{equation}
Shen et al.\ have proven that the total Berry charge is the same for these four possibilities \cite{shen_topological_2017}.
Thus, here we compute the local Berry curvature, $\boldsymbol{\Omega}(\mathbf{k}) = \nabla_{\mathbf{k}} \times \mathbf{A}(\mathbf{k})$
for the $\lambda_+$ band of non-Hermitian system in Eq.\ (\ref{eq:nhw}) as,
\begin{align}
  \Omega_{\rho}^{LR} =& \left(\frac{n k_\rho^{n-1}}{2 (\lambda-\omega_0)^3} \right) \left(k_\rho^n + i \tau_x \cos(n \phi) - i \tau_y \sin(n \phi) \right), \\
  \Omega_{\phi}^{LR} =& \left(\frac{-i n k_\rho^{n-1}}{2 (\lambda-\omega_0)^3} \right) \left(\tau_x \cos(n \phi) + \tau_y \sin(n \phi) \right), \\
  \Omega_z^{LR} =& \left(\frac{n^2 k_\rho^{2(n-1)}}{2 (\lambda-\omega_0)^3} \right)(k_z + i \tau_z),
\end{align}
for which we define the Berry connection as $\mathbf{A}^{LR} = i \langle \psi(\mathbf{k})^{L} | \nabla_{\mathbf{k}} | \psi(\mathbf{k})^{R} \rangle$.
Upon integrating the Berry curvature on a closed surface containing the exceptional contour, one can
analytically demonstrate that the total Berry charge is still real and quantized,
\begin{equation}
  \gamma = \int_S \boldsymbol{\Omega}^{LR}(\mathbf{k}) \cdot d\mathbf{S} = n,
\end{equation}
for which the full proof is given in Appendix \ref{app:a}. If the Berry curvature is integrated on a closed surface which does not enclose
any portion of the exceptional contour, the resulting charge is zero \cite{note1}.
Thus, as gain and loss are added to
any topologically charged system, the topological charge is preserved on the
exceptional contour which forms from the original topologically charged degeneracy.

\begin{figure}[t!]
  \centering
  \includegraphics[width=0.98\linewidth]{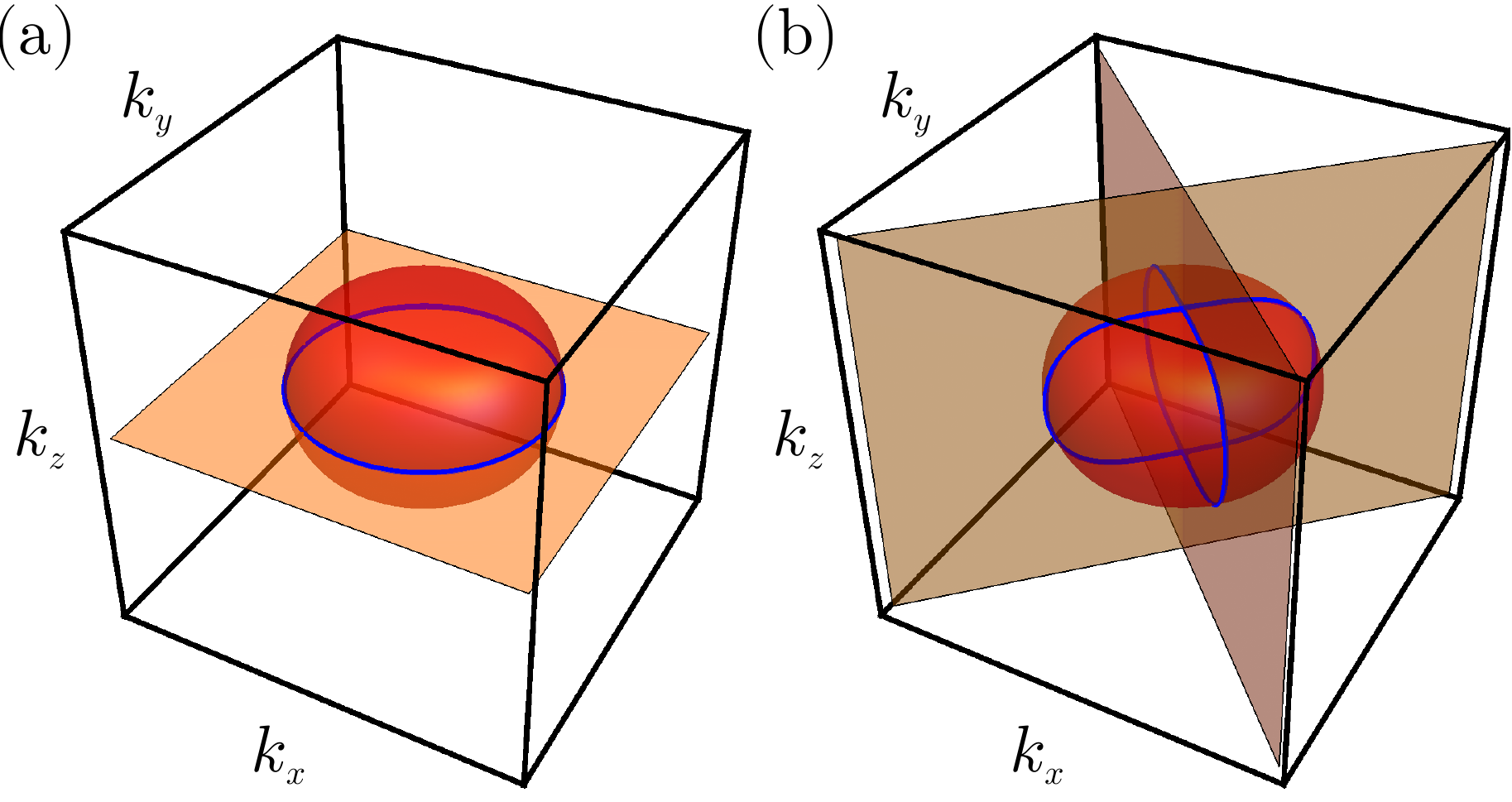}
  \caption{(a) Exceptional contour associated with a charge-1 Weyl point under non-Hermitian perturbation,
    as described by Eq.\ (\ref{eq:nhw}) with $n=1$ and $\boldsymbol{\tau} = (0,0,\tau_z)$.
    (b) Exceptional contour associated with a charge-2 Weyl point under non-Hermitian perturbation,
    as described by Eq.\ (\ref{eq:nhw}) with $n=2$ and $\boldsymbol{\tau} = (\tau_x,0,0)$.
    In both panels, the two surfaces plotted are Eq.\ (\ref{eq:5}) (red) and Eq.\ (\ref{eq:6}) (orange),
    and their intersection defines the exceptional contour (blue). }
  \label{fig:1}
\end{figure}

Previous studies on the effects of adding non-Hermitian material to otherwise Hermitian systems
with isolated degeneracies have focused on Dirac points and charge-1 Weyl points, for which
the resulting exceptional contour in both cases is a ring \cite{szameit_pt-symmetry_2011,zhen_spawning_2015,xu_weyl_2017,shen_topological_2017}, shown in Fig.\ \ref{fig:1}(a). However, the general
requirements for the formation of the exceptional contour, Eqs.\ (\ref{eq:5}) and (\ref{eq:6}),
do not necessitate this outcome. For example, in a non-Hermitian system with an underlying charge-2 Weyl
point described by Eq.\ (\ref{eq:nhw}) with $n=2$ and $\boldsymbol{\tau} = (\tau_x, 0, 0)$, the resulting
exceptional contour consists of two intersecting rings, and is shown in Fig.\ \ref{fig:1}(b). 

It is worth remarking on the difference between adding a Hermitian versus non-Hermitian perturbation
to a Hamiltonian containing a charge-$n$ Weyl point, Eq.\ (\ref{eq:Hherm}). The consequence of adding
a general Hermitian perturbation to this Hamiltonian is to break the symmetry protecting the charge-$n$ Weyl point,
breaking up the single Weyl point into $n$ charge-1 Weyl points. In contrast, upon adding a general anti-Hermitian
perturbation of the form considered in Eq.\ (\ref{eq:nhw}), the charge-$n$ Weyl point transforms into a single
exceptional contour with charge $n$, rather than forming $n$ charge-1 exceptional contours, a result which
is guaranteed by the form of Eqs.\ (\ref{eq:5}) and (\ref{eq:6}). Moreover, a
non-Hermitian perturbation can be used to reconstruct a single charge-$n$ exceptional contour from $n$ charge-1
Weyl points which have been split apart by a Hermitian perturbation.

\subsection{A general criteria for forming exceptional contour in two-band systems \label{sec:2c}}

In Sec.\ \ref{sec:2a}, we derived the two criteria for finding an exceptional
contour when a fixed amount of gain and loss is added to an underlying Hermitian
system with a charge-$n$ Weyl point in Eqs.\ (\ref{eq:5}) and (\ref{eq:6}).
These two conditions can be generalized to a broader class of non-Hermitian Hamiltonians
in which the added gain and loss is wavevector dependent, $\boldsymbol{\tau}(\mathbf{k})$.
Consider a generic $2 \times 2$ non-Hermitian Hamiltonian,
\begin{equation}
  H(\mathbf{k}) = f_x(\mathbf{k}) \sigma_x + f_y(\mathbf{k}) \sigma_y + f_z(\mathbf{k}) \sigma_z + \omega_0 I, \label{eq:gen}
\end{equation}
in which the functions $f_i(\mathbf{k})$ are complex, smooth functions which describe
the band structure. The upper and lower branches of the eigenvalues of this system can be written as
\begin{align}
  \lambda_\pm - \omega_0 &= \pm \sqrt{f_x^2(\mathbf{k}) + f_y^2(\mathbf{k}) + f_z^2(\mathbf{k})} \notag \\
  &= \pm \sqrt{\det[H-\omega_0 I]},
\end{align}
so that the general requirements for finding $\lambda_+ = \lambda_-$ are
\begin{align}
  \re[\det[H-\omega_0 I]] = 0, \label{eq:cond1} \\
  \im[\det[H-\omega_0 I]] = 0. \label{eq:cond2}
\end{align}
Again, as there are only two constraints but three degrees of freedom,
and thus absent some additional system symmetry which allows higher dimensional surfaces to form,
we expect to only find one-dimensional contours (i.e.\ lines) of exceptional points.

Moreover these lines in parameter space where $\lambda_+ = \lambda_-$ are comprised entirely
of exceptional points. To prove this, we first assume that our system is not at a point where
$H(\mathbf{k}_0)-\omega_0 I = 0$. Then, we solve for the left and right eigenvectors
of the generic non-Hermitian system in Eq.\ (\ref{eq:gen}),
\begin{align}
  |\psi_\pm^R \rangle &= \left( f_x(\mathbf{k}) - i f_y(\mathbf{k}), \; \lambda_\pm - \omega_0 - f_z(\mathbf{k}) \right)^T, \\
  \langle \psi_\pm^L| &= \left( f_x(\mathbf{k}) + i f_y(\mathbf{k}), \; \lambda_\pm - \omega_0 - f_z(\mathbf{k}) \right),
\end{align}
and then calculate their inner product,
\begin{equation}
  \langle \psi_\pm^L| \psi_\pm^R \rangle = 2 \det[H-\omega_0 I] \mp 2 f_z(\mathbf{k}) \sqrt{\det[H-\omega_0 I]}.
\end{equation}
As the condition for $\lambda_+ = \lambda_-$ is $\det[H(\mathbf{k}_0)-\omega_0 I] = 0$, at every point
where these two eigenvalues are equal the eigenvectors are self-orthogonal, $\langle \psi_\pm^L| \psi_\pm^R \rangle = 0$,
and so these lines in parameter space where the two eigenvalues are equal are guaranteed
to be lines of exceptional points.

\subsection{Topological phase transition induced by exceptional contour merging \label{sec:2d}}

\begin{figure}[t!]
  \centering
  \includegraphics[width=0.98\linewidth]{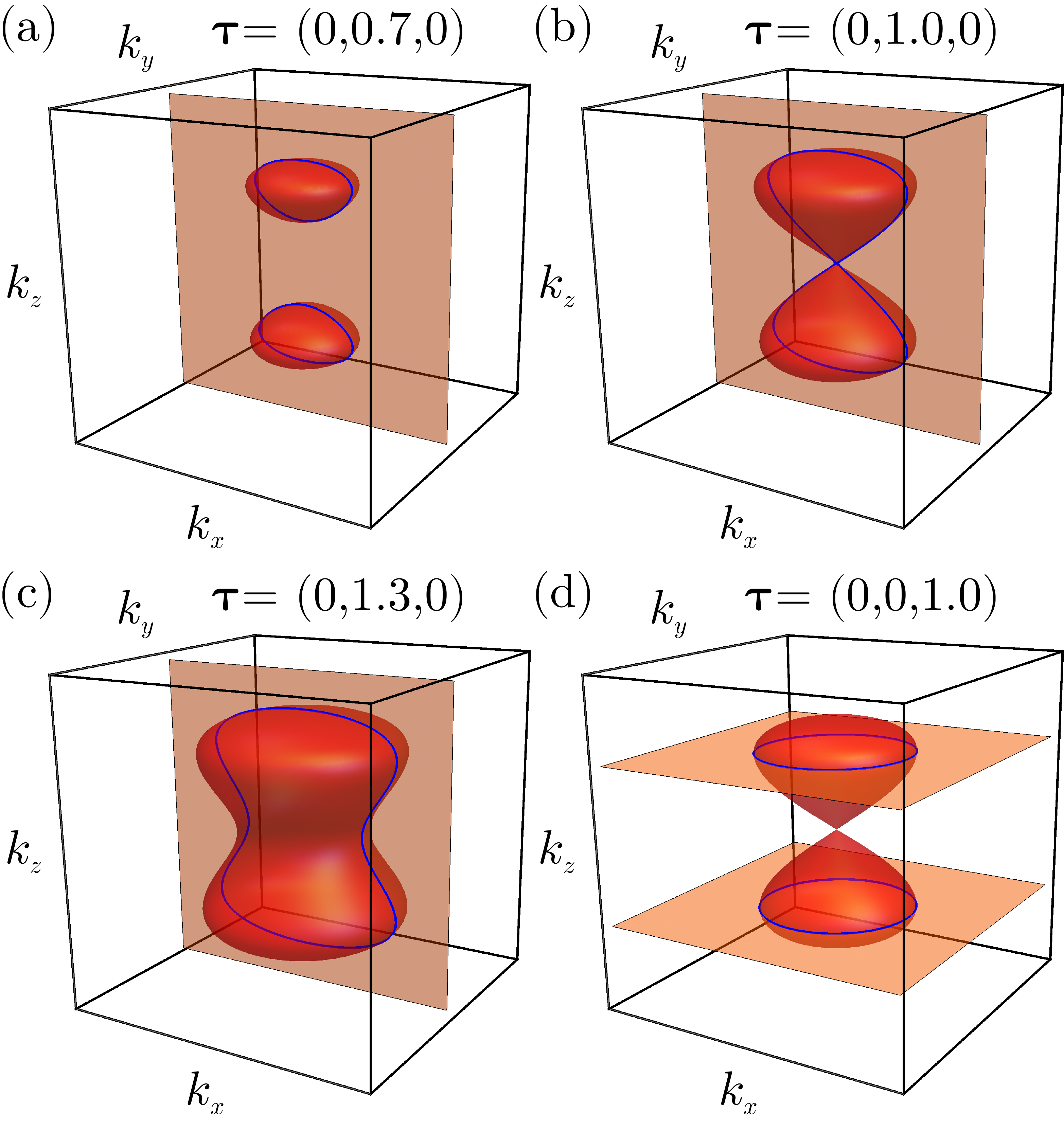}
  \caption{Exceptional contours for the Hamitonian of Eq.\ (\ref{eq:twoEC}) with $k_0 = 1$. In all
    panels, the two surfaces plotted are Eq.\ (\ref{eq:5}) (red) and Eq.\ (\ref{eq:6}) (orange), and their
    intersection defines the exceptional contour (blue).
    (a) $\boldsymbol{\tau} = (0,0.7,0)$. (b) $\boldsymbol{\tau} = (0,1.0,0)$.
    At this value of $\boldsymbol{\tau}$, the exceptional contours touch, the Berry flux dissipates,
    and the exceptional coutour become uncharged. 
    (c) $\boldsymbol{\tau} = (0,1.3,0)$.
    (d) $\boldsymbol{\tau} = (0,0,1.0)$. Regardless of the strength of $\tau_z$, the exceptional
    contours never merge for $\boldsymbol{\tau} = \tau_z$.}
  \label{fig:merging}
\end{figure}

In Hermitian systems with two bands, the only way to remove a Weyl point from the band structure is for it to combine with an
oppositely charged Weyl point, after which the two bands develop a band gap. However, the addition of non-Hermitian
terms to a system's Hamiltonian provides a different mechanism by which topological charge can be dissipated,
when the exceptional contours with opposite topological charge merge to become a single, uncharged exceptional contour.
To give a simple example of this phenomena, consider the system,
\begin{equation}
  H = k_x \sigma_x + k_y \sigma_y + (k_z^2 - k_0^2)\sigma_z + i \boldsymbol{\tau} \cdot \boldsymbol{\sigma}, \label{eq:twoEC}
\end{equation}
which contains two Weyl points with opposite charge at $k_z = \pm k_0$ for $\boldsymbol{\tau} = 0$.
As gain and loss is added to the system with $\boldsymbol{\tau} = (0,\tau_y,0)$, each of these charge-1 Weyl points becomes a charge-1
Weyl exceptional contour, as can be seen in Fig.\ \ref{fig:merging}(a). But, as the strength of the gain and
loss is increased, the two Weyl exceptional contours eventually touch at the threshold $\tau_y = k_0^2$, as shown in Fig.\ \ref{fig:merging}(b).
When this happens, it becomes impossible to draw a surface over which to calculate the Berry charge that
only contains a single exceptional contour, and as such the topological charge contained on each of the Weyl
exceptional rings is dissipated, leaving a single uncharged exceptional contour, shown in Fig.\ \ref{fig:merging}(c).
Interestingly, in this example, the topological charge can dissipate without opening a gap.

The existance of such a threshold is dependent upon the form of the non-Hermitian Hamiltonian.
To illustrate this, in Fig.\ \ref{fig:merging}(d) we show the same system except with $\boldsymbol{\tau} = (0,0,\tau_z)$.
As can be seen, the surfaces corresponding to the second criteria for finding the exceptional contour, Eq.\ (\ref{eq:cond2}),
run parallel to one another, regardless of the value of $\tau_z$. Thus, the two exceptional contours for this system
for this distribution of gain and loss will never touch for any strength of the added gain and loss and hence
will retain their topological charge.

\subsection{Multiple exceptional contours \label{sec:2e}}

So far, we have focused on adding a constant strength of gain and loss to particular lattice sites.
In this section we briefly explore some of the additional complexities which can arise
if the non-Hermitian perturbation to the system is dependent upon the wavevector, i.e.\ $\boldsymbol{\tau} = \boldsymbol{\tau}(\mathbf{k})$.
In this case, it is possible to find that the intersection of Eqs. (\ref{eq:cond1}) and (\ref{eq:cond2})
yields multiple separate exceptional contours.

\begin{figure}[t]
  \centering
  \includegraphics[width=0.98\linewidth]{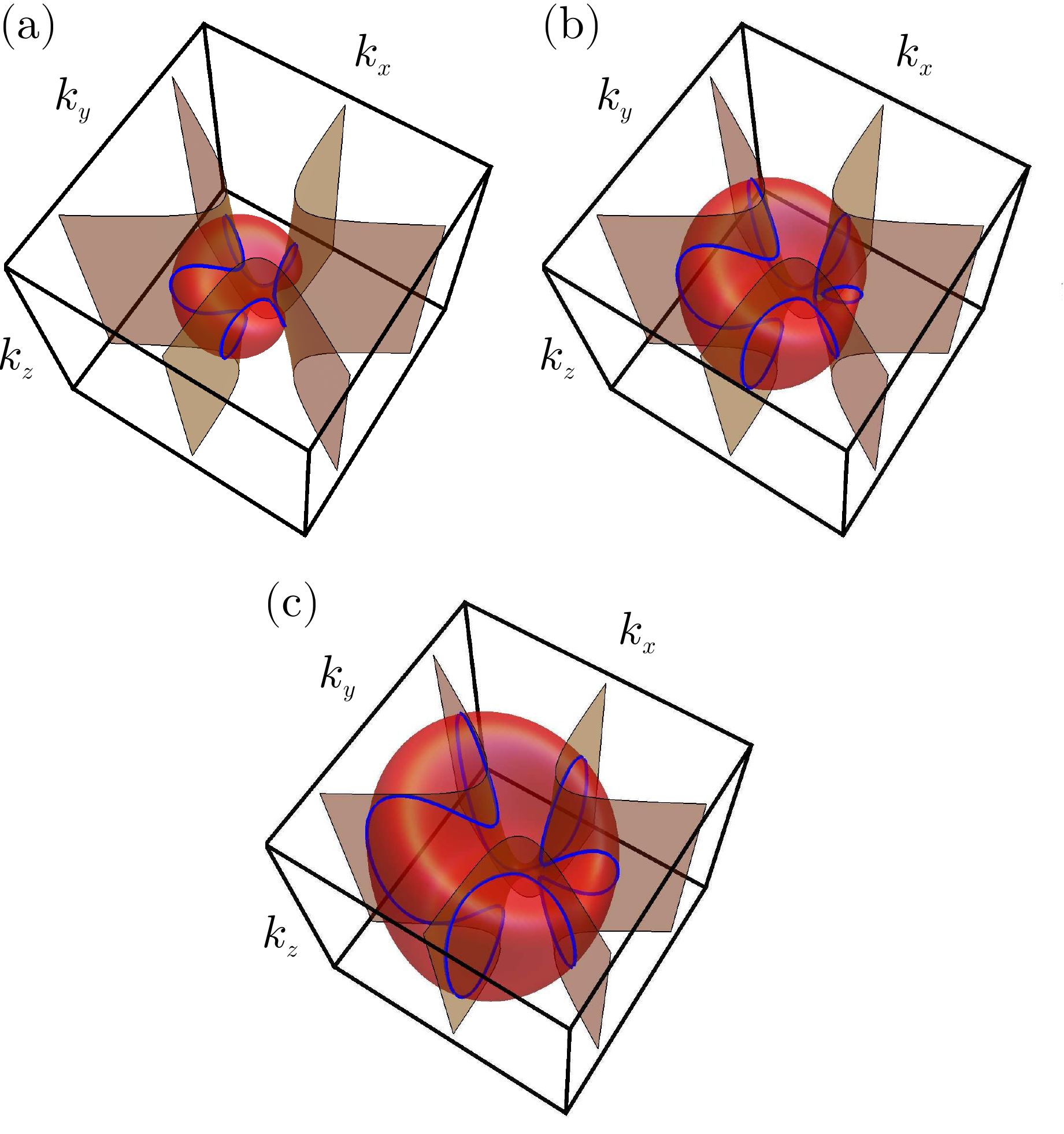}
  \caption{Exceptional contour associated with a charge-2 Weyl point with wavevector-dependent gain and loss
    added, as described by Eq.\ (\ref{eq:nhw}) with $n=2$ and
    $\boldsymbol{\tau}(\mathbf{k}) = \tau_0 (-4 k_x + \sqrt{12}, -4 k_y + 1, -1/\sqrt{2})$, with
    $\tau_0 = 0.7$ (a), $\tau_0 = 1.0$ (b), and $\tau_0 = 1.3$ (c).
    The two surfaces plotted are Eq.\ (\ref{eq:cond1}) (red) and Eq.\ (\ref{eq:cond2}) (orange),
    and their intersection defines the exceptional contour (blue). As can be seen, a portion of the exceptional contour splits off
    for $\tau_0 \sim 1$, yielding two independent exceptional contours. When this happens, the smaller exceptional contour does
    not carry a Berry charge, while the larger exceptional contour still carries a Berry charge of $2$.}
  \label{fig:EC2}
\end{figure}

For example, consider the charge-2 system described by Eq.\ (\ref{eq:nhw}) with $\boldsymbol{\tau}(\mathbf{k}) = \tau_0 (-4 k_x + \sqrt{12}, -4 k_y + 1, -1/\sqrt{2})$,
in which $\tau_0$ provides an overall scaling to the strength of the gain and loss.
This system exhibits three distinct phases depending on the value of $\tau_0$. First, for small values
of $\tau_0$, the system possesses a single exceptional contour with a Berry charge of 2, similar to the
examples considered in previous sections and shown in Fig.\ \ref{fig:EC2}(a). Next, as the strength of the gain and loss is increased beyond
a threshold value, $\tau_0 \ge \tau_{2c}$, a second exceptional contour appears which is separate from the original
Weyl exceptional contour, shown in Fig.\ \ref{fig:EC2}(b). This second exceptional contour does not possess
a Berry charge. Finally, as $\tau_0$ is increased past a second threshold value, $\tau_0 \ge \tau_{1c}$,
the two exceptional contours merge, and form a single exceptional contour with charge 2, shown in Fig.\ \ref{fig:EC2}(c).
This example also illustrates that even when the added wavevector-dependent gain and loss breaks some of the symmetries
of the underlying Hermitian system, the entire quantized Berry charge of the system can still be found on a single exceptional contour.
In this example, when $\tau_0 = 0$ the charge-2 Weyl point in $H$ is protected by $C_{4}$ symmetry,
but this is not true for $\tau_0 \ne 0$, and the resulting exceptional contour
still has a Berry charge of $2$, rather than splitting into two charge-1 exceptional contours.

Note that in choosing to add a wavevector dependent non-Hermitian perturbation to the system, we stipulate that as each $k_i \rightarrow \pm \infty$,
the Hermitian portion of the Hamiltonian should dominate the behavior of the system, so that the non-Hermitian addition can still
be considered a perturbation on the underlying Hermitian system. To satisfy this criteria and still be able to add a
wavevector dependent non-Hermitian term to the system, the minimum charge of the Weyl point in the underlying Hermitian system is $2$.


\section{Photonic realizations of Weyl exceptional contours \label{sec:3}}

In the previous section, using a two-band effective Hamiltonian, we describe a number of interesting effects,
when a non-Hermitian perturbation is added to a topologically non-trivial Hermitian Hamiltonian
supporting Weyl points. These effects include the formation of topologically charged exceptional
contours with unusual shapes and the dissipation of topological charge without opening a band gap.
In this section we will show that these novel physical effects can be realized in realistic photonic systems.

\begin{figure}[t]
  \centering
  \includegraphics[width=0.70\linewidth]{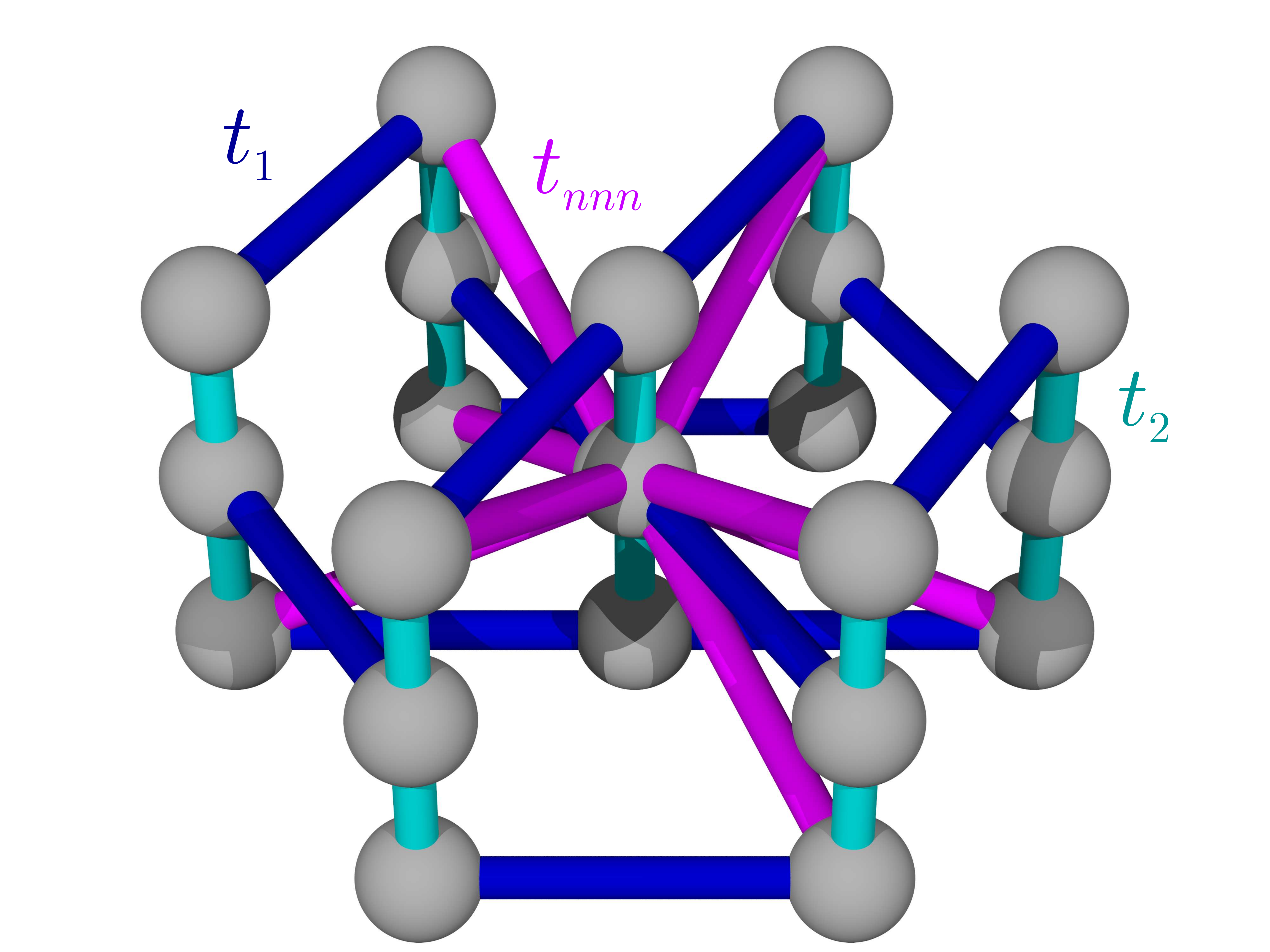}
  \caption{Schematic of the tight binding model for a metallic chiral woodpile photonic crystal \cite{chang_multiple_2017}. The nearest neighbor
    bonds are shown in blue ($t_{n1}$) and cyan ($t_{n2}$), while the next nearest neighbor bonds connected to the center
    site are shown in purple ($t_{nnn}$).}
  \label{fig:charge2schem}
\end{figure}

To observe a non-ring Weyl exceptional contour, we use a metallic chiral woodpile photonic crystal designed to
operate in the terahertz frequency band,
for which the underlying Hermitian system can possess charge-2 Weyl points.
The details of the structure can be found in Ref.\ \cite{chang_multiple_2017}. For the purposes of this paper,
following \cite{chang_multiple_2017}, we note that the properties of this structure are well described by the tight-binding model as
shown in Fig.\ \ref{fig:charge2schem}. The model consists of layers of a hexagonal lattice directly on top of each other.
The lattice sites (grey spheres) are  
connected by nearest-neighbor hopping, intra-layer hopping along only one of the lattice vector directions with strength $t_{n1}$,
shown as blue bonds in Fig.\ \ref{fig:charge2schem}, and inter-layer
hopping with strength $t_{n2}$, shown as cyan bonds in Fig.\ \ref{fig:charge2schem}. In addition, there is a set of
next-nearest-neighbor hopping terms, with strength $t_{nnn}$, and all of these bonds connected to the
central lattice site in Fig.\ \ref{fig:charge2schem} are shown as purple bonds. The tight binding
Hamiltonian for this system is then
\begin{multline}
  H = \sum_{i,k} \varepsilon_k a_{i,k}^\dagger a_{i,k} + \sum_{\langle i,j \rangle} t_{n1} a_{i,k}^\dagger a_{j,k} + \textrm{c.c.} \\
  + \sum_{i,k} t_{n2} a_{i,k}^\dagger a_{i,k+1} + \textrm{c.c.} + \sum_{\langle \langle i,j \rangle \rangle} t_{nnn} a_{i,k}^\dagger a_{j,k+1} + \textrm{c.c.}, \label{eq:tbm1}
\end{multline}
in which $a_{i,k}$ and $a_{i,k}^\dagger$ are the annihilation and creation operators at the $i$th lattice site within
the $k$th lattice layer, $\langle i,j \rangle$ denotes a pair of intra-layer nearest neighbors, and $\langle \langle i, j \rangle \rangle$
denotes a pair of inter-layer next nearest neighbor couplings. If we set $t_{n1} = 1$, $t_{n2} = -1$, and $t_{nnn} = 0.2$, this three-band model contains a charge-2 Weyl
point between two bands at $\Gamma$, and a set of charge-1 Weyl points between the same bands at $K$, such that the
total Berry charge in the $k_z = 0$ plane is zero. Likewise, a second charge-2 Weyl point can be found at $A$ alongside
charge-1 Weyl points at $H$, again such that there is no net Berry charge in the $k_z = \pm \pi/2$ plane \cite{chang_multiple_2017}.

\begin{figure}[t]
  \centering
  \includegraphics[width=0.98\linewidth]{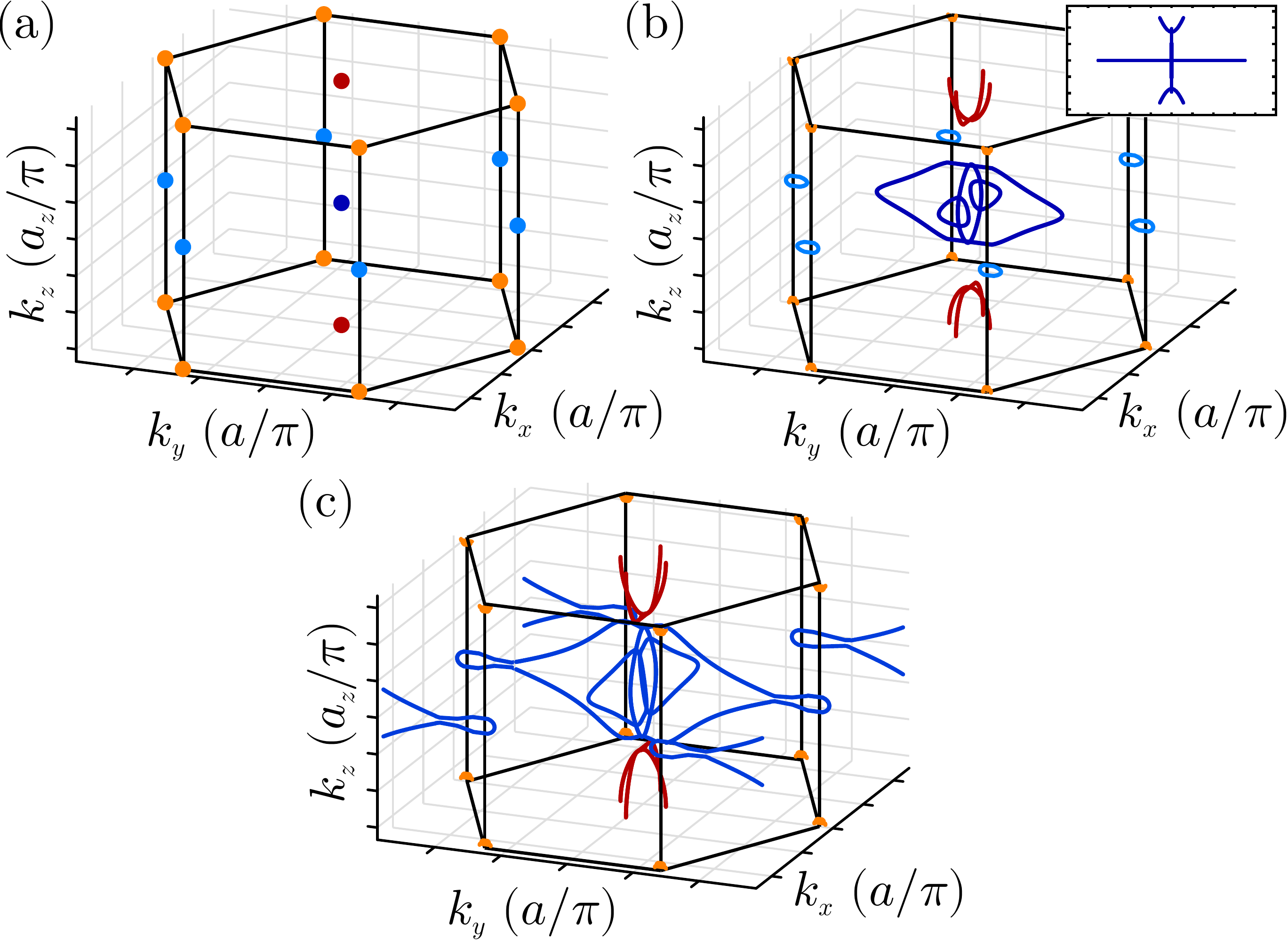}
  \caption{Plot of the (a) Weyl points and (b),(c) Weyl exceptional contours of the tight binding
    model given in Eq.\ (\ref{eq:tbm1}) with hopping strengths $t_{n1} = 1$, $t_{n2} = -1$, and $t_{nnn} = 0.2$.
    In (b), gain and loss has been added to two of the three lattice sites in the unit cell, with $\varepsilon_1 = \varepsilon_3^* = 0.2i$,
    and $\varepsilon_2 = 0$. The two charge-2 Weyl points and exceptional contours at $\Gamma$ and $A$ are shown in dark blue and
    dark red respectively, while the numerous charge-1 Weyl points and exceptional contours at $K$ and $H$ are shown in light blue and orange.
    A top-down view of the charge-2 Weyl exceptional contour at $\Gamma$ is shown in the inset.
    In (c), sufficient gain and loss has been added, $\varepsilon_1 = \varepsilon_3^* = 0.29i$, to cause the charge-2 Weyl exceptional contour at $\Gamma$ to merge with
    two of the charge-1 Weyl exceptional contours at $K$, forming a single uncharged exceptional contour.}
  \label{fig:charge2}
\end{figure}

To break Hermiticity in the system, we
allow the on-site energies in each lattice layer to be complex, $\varepsilon_k \in \mathbb{C}$, which corresponds
to adding gain or loss to that layer of the photonic crystal. The resulting Weyl exceptional contours in this
system are shown in Fig.\ \ref{fig:charge2}, in which the exceptional contour stemming from the charge-2 Weyl point
at $A$ resembles two intersecting rings, which is similar to the shape of the charge-2 Weyl exceptional contour considered in Fig.\ \ref{fig:1}(b).
The charge-2 Weyl exceptional contour at $\Gamma$ consists of four interlocking rings, the larger two rings are centered at $\Gamma$, while
the smaller two rings are situated opposite one another in $k_x$, demonstrating that additional complexities are possible
in realistic systems beyond the simple systems considered in Sec.\ \ref{sec:theory}. A top-down view of the charge-2 Weyl exceptional contour
at $\Gamma$ is shown in the inset of Fig.\ \ref{fig:charge2}(b).
Additional charge-1 Weyl exceptional rings are seen at $K$ and $H$. The quantization
of the Berry charge for all of these Weyl exceptional contours is confirmed numerically.

As the gain and loss in this system is further increased, the topologically charged exceptional contours
begin to merge together to form topologically trivial exceptional contours, in agreement with
the systems considered in Sec.\ \ref{sec:2d}. In Fig.\ \ref{fig:charge2}(c), the charge-2 Weyl exceptional
contour at $\Gamma$ has merged with two of the charge-1 Weyl exceptional contours at $K$ to form a single
uncharged exceptional contour.

\section{Chiral edge modes \label{sec:haldane}}

One of the most important properties of Hermitian Weyl semi-metals are their surface states, which allow
for one-way transport. As such, it is of critical importance to understand whether surface
states persist in non-Hermitian topological systems, and if so how they are effected by the presence of
the gain and loss. Here we study a
non-Hermitian extension of a three-dimensional
Hermitian system exhibiting a charge-1 Weyl point \cite{haldane_model_1988}, 
and demonstrate that surface states of this system are preserved outside of the Weyl exceptional contour.

\begin{figure}[t]
  \centering
  \includegraphics[width=0.70\linewidth]{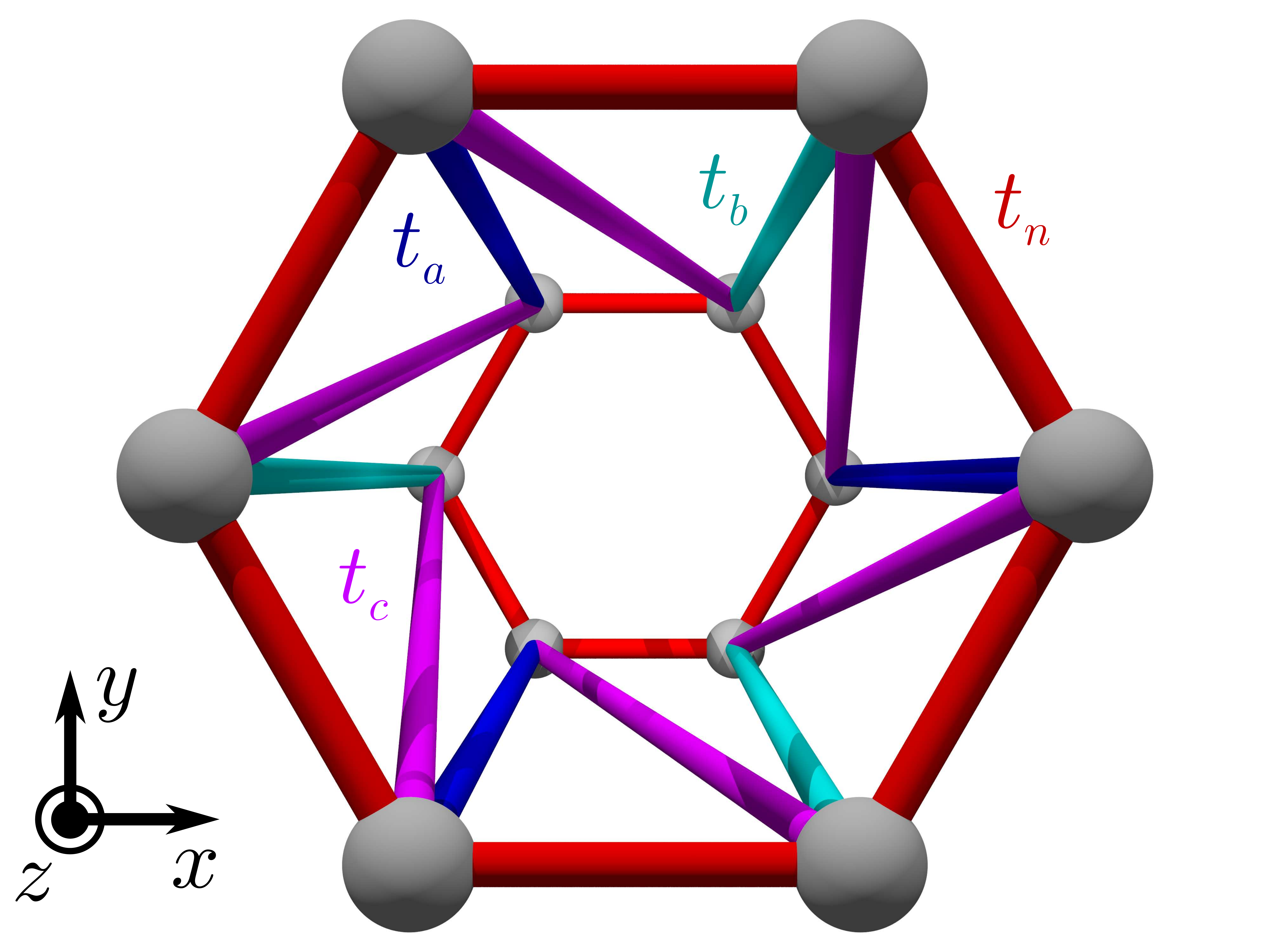}
  \caption{Schematic of the tight binding model for the Hamiltonian given in Eq.\ (\ref{eq:hermH}), which can be
    realized in both acoustic \cite{xiao_synthetic_2015} and photonic \cite{lin_photonic_2016} systems. The intra-layer
    bonds with strength $t_n$ are red, the direct inter-layer bonds for the A and B sites are $t_a$ (blue) and $t_b$ (cyan), respectively,
    and the inter-layer next nearest neighbor bonds, $t_c$, are shown in purple.}
  \label{fig:surfSchem}
\end{figure}

The non-Hermitian three-dimensional model we consider consists of layers of a stacked honeycomb lattice, in which
all layers are directly stacked on top of each other, as shown in Fig.\ \ref{fig:surfSchem}. The Hamiltonian
reads,
\begin{multline}
  H_\perp = \sum_{i,m} \varepsilon_a a_{i,m}^\dagger a_{i,m} + \varepsilon_b b_{i,m}^\dagger b_{i,m} \\
  + \sum_{\langle i,j \rangle,m} t_n \left(a_{i,m}^\dagger b_{j,m} + b_{j,m}^\dagger a_{i,m} \right).
\end{multline}
Here, $a_{i,m},b_{i,m}$ are the annihilation operators at the $i$th site on the $m$th layer 
of the two sub-lattices of the honeycomb crystal. The on-site energies of the two sub-lattices
are $\varepsilon_{a,b}$, the in-plane coupling strength is $t_n$, and $\langle i,j \rangle$ represents a pair of nearest neighbors.
In addition,
there are two forms of out-of-plane couplings, direct coupling between neighboring layers,
\begin{equation}
H_{D} = \sum_{i,m} t_a a_{i,m+1}^\dagger a_{i,m} + t_b b_{i,m+1}^\dagger b_{i,m} + \textrm{c.c.},
\end{equation}
and coupling between next-nearest neighbors on adjacent layers,
\begin{equation}
H_{NNN} = \sum_{\langle \langle i,j \rangle \rangle,m} t_c \left( a_{i,m+1}^\dagger a_{j,m} + b_{i,m+1}^\dagger b_{j,m} \right) + \textrm{c.c.},
\end{equation}
in which $\langle \langle i,j \rangle \rangle$ represents the possible pairs of inter-layer next nearest neighbors.
These three different couplings are shown schematically in Fig.\ \ref{fig:surfSchem}. Together,
the total system
\begin{equation}
H_0 = H_\perp + H_D + H_{NNN}, \label{eq:hermH}
\end{equation}
contains a set of Weyl points, whose positions depend upon the choice of the
coupling strength parameters \cite{xiao_synthetic_2015}.
This Hamiltonian has been previous shown to be realizable either in acoustic systems \cite{xiao_synthetic_2015},
or in a photonic system using the concept of synthetic dimensions \cite{lin_photonic_2016}. 

\begin{figure}[t]
\centering
\includegraphics[width=0.98\linewidth]{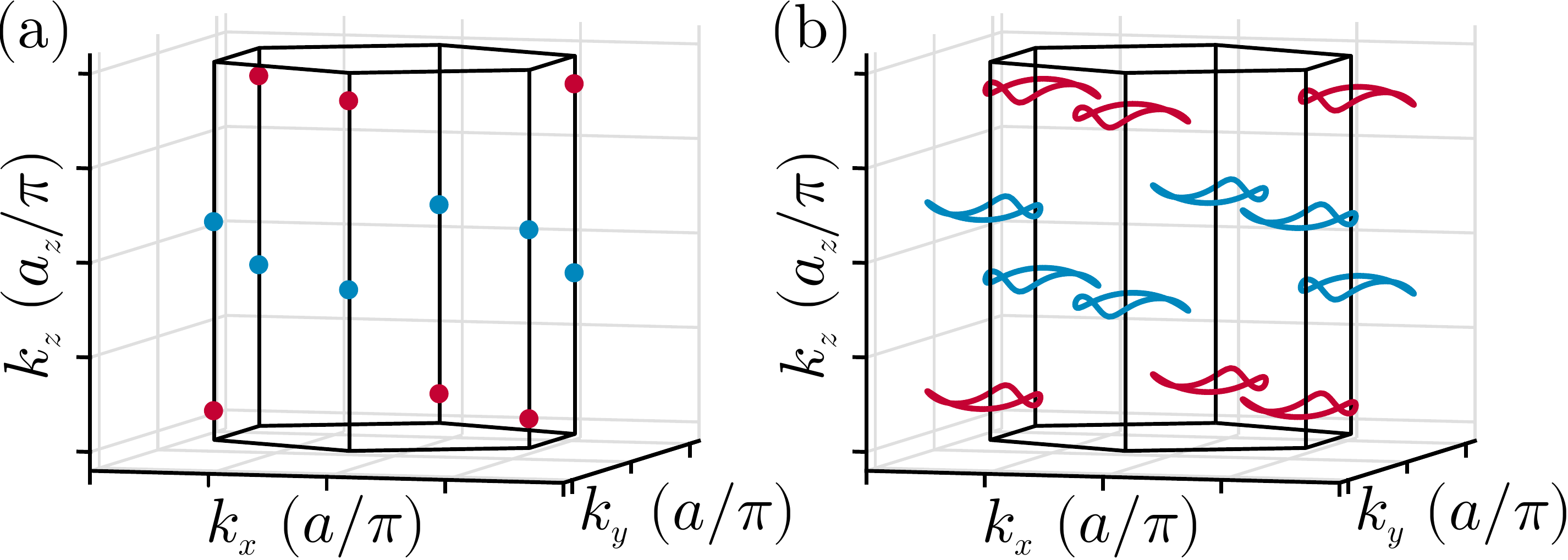}
\caption{Locations of Weyl points (a) and Weyl exceptional rings (b)
  for the Hamiltonian given in Eq.\ (\ref{eq:hermH}) with $t_a = -t_b = 4$, $t_c = 3$, and $t_n = 10$.
  In (b), gain and loss have been added to the A and B sites of the lattice, with $\varepsilon_a = \varepsilon_b^* = 9i$.}
  \label{fig:surfContours}
\end{figure}

\begin{figure*}[ht!]
\centering
\includegraphics[width=0.95\linewidth]{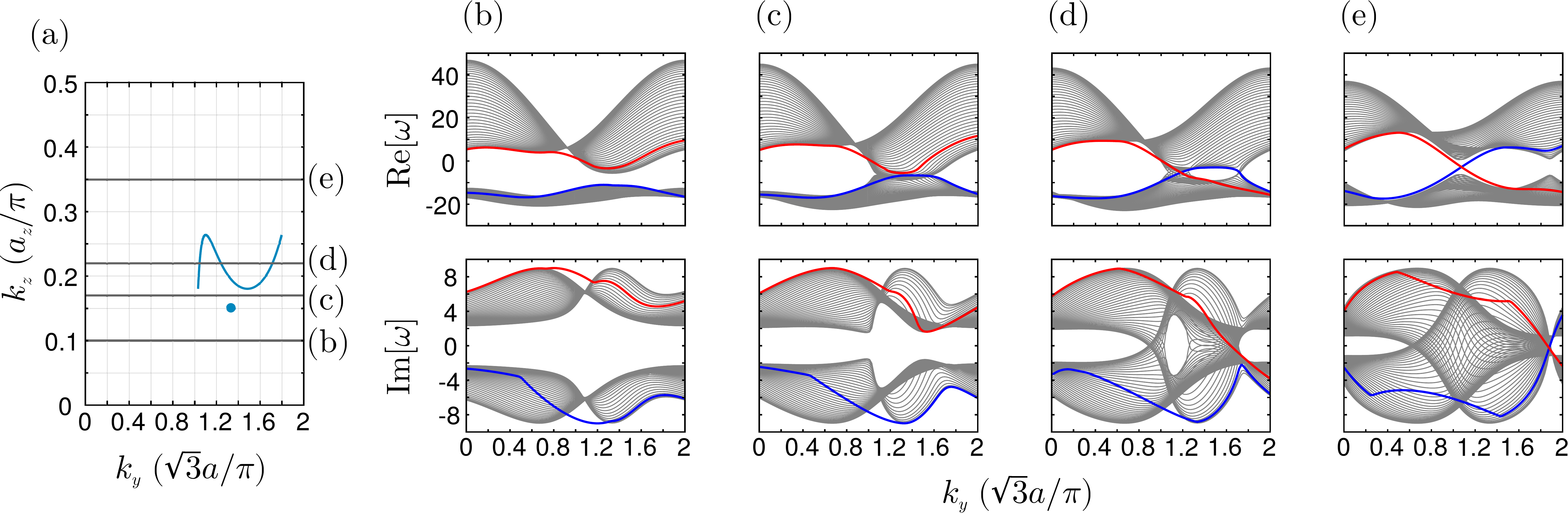}
\caption{(a) Projection of the Weyl point (solid circle) of the underlying Hermitian system given in Eq.\ (\ref{eq:hermH}) and the exceptional
  contour of the full non-Hermitian system in the $k_y$-$k_z$ plane,
  with $t_a = -t_b = 4$, $t_c = 3$, $t_n = 10$, and $\varepsilon_a = \varepsilon_b^* = 9i$. The four gray lines denote the value of
  $k_z$ chosen to calculate the band structure in plots (b)-(e). (b) $k_z$ is below the
  exceptional contour and the upper and lower bands are separated by a topologically trivial gap, so
  no topologically protected surface states are seen.
  (c) As $k_z$ approaches the exceptional contour,
  the upper and lower bands overlap within the region where the exceptional contour forms.
  (d) $k_z$ is in the middle of the exceptional contour, and some of the lower bands cross over to become
  upper bands within the exceptional contour, and vise versa.
  (e) $k_z$ is greater than the entire exceptional contour and thus a topologically non-trivial gap opens
  between the upper and lower bands.
  In (b)-(d), the blue and red states correspond to surface states which are
  not topologically protected, while in (e) these two states correspond to the topologically protected surface states.}
\label{fig:surf}
\end{figure*}

To break Hermiticity in this system, gain and loss are added to the two on-site energies, and
are chosen such that $\varepsilon_b = \varepsilon_a^*$ to enforce that an equal amount of gain and loss is added to the
two lattice sites. In doing so, each of the charge-1 Weyl points in the underlying Hermitian system
turns into a Weyl exceptional ring, as shown in Fig.\ \ref{fig:surfContours}. As the strength of the gain
and loss is increased, these exceptional rings are no longer centered on the original Weyl points
in the system.

The consequences of these changes can be seen in the surface states of the surface formed by cutting
through all of the layers along a zig-zag edge at the same in-plane location, i.e.\ the cut is made
parallel to the $z$ axis as defined in Fig.\ \ref{fig:surfSchem}.
The band structures for a semi-infinite system containing two such surfaces along the $x$ axis
are shown in Fig.\ \ref{fig:surf} for a plane of constant $k_z$,
for four different choices of $k_z$. First, when $k_z$ is below
the exceptional contour, the upper and lower bands are separated by a topologically trivial gap, and no
topologically protected surface states are seen (Fig.\ \ref{fig:surf}(b)). As $k_z$ approaches the
exceptional contour, the upper and lower bands begin to overlap within the region where the
exceptional contour forms (Fig.\ \ref{fig:surf}(c)). When $k_z$ is within the the exceptional contour,
some of the lower bulk bands cross over through the exceptional contour to the upper band, and
vise versa, as seen in Fig.\ \ref{fig:surf}(d). This cross-over of the bulk bands is the manifestation
of the intersecting sheets of the Riemann surface topology between the branch points which comprise the
exceptional contour, as exemplified in Figs.\ \ref{fig:0}(b) and (c). Finally, when $k_z$ is increased past the entire
exceptional contour, a topologically non-trivial gap opens between the upper and
lower bands. Within this gap there are topologically protected surface states. These states have non-zero imaginary
components in their eigenvalues, indicating the novel possibility of one-way amplification or dissipation in these systems.

\section{Conclusion \label{sec:4}}

In conclusion, we have provided a systematic study of the effects of introducing non-Hermiticity to
topologically charged systems. In doing so, we have proven that a Weyl point with arbitrary charge in
a Hermitian system transforms into a one-dimensional exceptional contour on which the topological charge
is preserved. Moreover, we have shown that the addition of gain and loss may result in a new class of
topological phase transition, throuh which the topological charge can dissipate without opening a gap.
Our results highlight significant opportunities for exploring topological physics in non-Hermitian systems. 

\appendix
\section{Proof of the quantized Berry charge \label{app:a}}

In this section, we provide an analytic proof that the exceptional contour in the non-Hermitian Hamiltonian given
in Eq.\ (\ref{eq:nhw}) still possesses a quantized topological charge identical to that of the charge of the Weyl point
in the underlying Hermitian system when $\boldsymbol{\tau} = 0$. To begin, we choose a cylinder in wavevector space as
our surface of integration, with radius $R$ and height $2Z$ centered at the origin. Additionally, we require that
$R^{2n} + Z^2 > \boldsymbol{\tau}^2$, so that no branch cuts intersect the surface.
The total topological charge of the system is then given by,
\begin{multline}
  \gamma = \int_0^{2 \pi} \left[ \int_0^R k_\rho \Omega_z^{LR}|_{k_z = Z} dk_\rho + R \int_{-Z}^Z \Omega_\rho^{LR}|_{k_\rho = R} dk_z  \right. \\
    \left. - \int_0^R k_\rho \Omega_z^{LR}|_{k_z = -Z} dk_\rho \right] d\phi,
\end{multline}
i.e.\ the integrals are over the top base, the side, and bottom base of the closed cylinder, respectively. Note that
the minus sign in the third term comes from $d\mathbf{S} = -\hat{z}$ on the bottom base.
The integrals over $k_z$ and $k_\rho$ can be directly evaluated as
\begin{widetext}
  \begin{multline}
    \int_0^R k_\rho \Omega_z^{LR}|_{k_z = Z} dk_\rho = \left(\frac{n}{2} \right) \left(\frac{Z+i \tau_z}{Z^2 - \boldsymbol{\tau}^2 + 2 i \tau_z Z + (\tau_x \cos(n \phi) - \tau_y \sin(n \phi))^2}\right) \\
    \times \left[ \sqrt{Z^2 - \boldsymbol{\tau}^2 + 2 i \tau_z Z}
      - \frac{Z^2 - \boldsymbol{\tau}^2 + 2 i \tau_z Z + iR^n (\tau_x \cos(n \phi) - \tau_y \sin(n \phi))}{\sqrt{R^{2n} + Z^2 - \boldsymbol{\tau}^2 +2i(\tau_x \cos(n \phi) - \tau_y \sin(n \phi)) + 2 i \tau_z Z}} \right], \label{eq:int1}
  \end{multline}
  \begin{multline}
    R \int_{-Z}^Z \Omega_\rho^{LR}|_{k_\rho = R} dk_z = \left(\frac{n}{2} \right) \left(\frac{R^{2n} + i R^n(\tau_x \cos(n \phi) - \tau_y \sin(n \phi)) }{R^{2n} -\tau_x^2 - \tau_y^2 + 2i R^n(\tau_x \cos(n \phi) - \tau_y \sin(n \phi))} \right) \\
    \times \left[ \frac{Z+i\tau_z}{\sqrt{R^{2n} + Z^2 - \boldsymbol{\tau}^2 +2i(\tau_x \cos(n \phi) - \tau_y \sin(n \phi)) + 2 i \tau_z Z}} + \frac{Z-i\tau_z}{\sqrt{R^{2n} + Z^2 - \boldsymbol{\tau}^2 +2i(\tau_x \cos(n \phi) - \tau_y \sin(n \phi)) - 2 i \tau_z Z}} \right], \label{eq:int2}
  \end{multline}
\end{widetext}
with the integral over the bottom base being identical to that of Eq.\ (\ref{eq:int1}), except with $Z \rightarrow -Z$.

Next, before evaluating the integral over $\phi$, we let $R \rightarrow \infty$, and keep only the leading order
non-zero terms. In doing so, Eq.\ (\ref{eq:int2}) becomes zero immediately. 
In addition, the denominator of the second term in the bracket of Eq.\ (\ref{eq:int1}) can be simply
replaced by $R^n$, and hence its contribution to the integral over $\phi$ vanishes.
The integration of Eq.\ (\ref{eq:int1}) over the azimuthal angle $\phi$ can now be evaluated using Cauchy's residue theorem,
which evaluates to $n/2$.
Thus, upon combining the contributions from both the top and bottom of the closed surface, we find that
$\gamma = n$, completing the proof.


\begin{acknowledgments}
This work was supported by an AFOSR MURI program (Grant
No.\ FA9550-12-1-0471), and an AFOSR project (Grant No.\ FA9550-16-1-0010).
\end{acknowledgments}


\begin{thebibliography}{72}
\bibitem [{not()}]{note1}%
  \BibitemOpen
  \href@noop {} {}\bibinfo {note} {As an exceptional point represents a
  singularity of the system where the two branches of $\lambda$ coalesce and
  there is only a single, self-orthogonal eigenvector, it remains an open
  question as to how to properly define the Berry connection and curvature at
  these points. Therefore, in this paper, we do not consider the case where the
  integration surface intersects with the contour.}\BibitemShut {Stop}%
\end{thebibliography}

%

\end{document}